\documentclass[12pt,reqno]{article}
\usepackage{amsmath}
\usepackage{graphicx}
\usepackage{color}
\usepackage{varioref}

\begin{document}

\title
\centerline{\bf\large Evolution of singlet structure functions from DGLAP equation at next-to-next-to-leading order at small-$x$}

\vspace{12pt}

\centerline{{\bf Mayuri Devee}$^{\rm 1}$ $^{\rm *}$, {\bf R. Baishya}$^{\rm 2}$ and {\bf J. K. Sarma }$^{\rm 1}$}
\centerline{$^{\rm 1}$ {HEP Laboratory, Department of Physics, Tezpur University, Tezpur, Assam, 784 028, India}} 
\centerline{$^{\rm 2}$ {Department of Physics, J. N. College, Boko-781123, Assam, India}} 
\centerline{$^{\rm *}${\it deveemayuri@gmail.com}}

\bigskip
\bigskip

\begin{abstract}
\leftskip1.0cm
\rightskip1.0cm
 A semi-numerical solution to Dokshitzer- Gribov-Lipatov-Altarelli-Parisi (DGLAP) evolution equations  at leading order (LO), next-to-leading order (NLO) and next-to-next-to-leading order (NNLO) in the small-$x$ limit is presented. Here we have used Taylor series expansion method to solve the evolution equations and, $t$- and $x$-evolutions of the singlet structure functions have been obtained with such solution. We have also calculated $t$- and $x$-evolutions of deuteron structure functions $F_{2}^{d}$, and the results are compared with the E665 data and NMC data. The results are also compared to those obtained by the fit to $F_{2}^{d}$  produced by the NNPDF collaboration based on the NMC and BCDMS data.



\end{abstract}

\section{Introduction}
\paragraph\ Structure functions in lepton-nucleon deep-inelastic scattering (DIS) are the established observables probing Quantum Chromodynamics (QCD), the theory of the strong interaction, and in particular the structure of the nucleon. The structure functions  provide unique information about the deep structure of the hadrons and most importantly, they form the backbone of our knowledge of the parton densities, which are indispensable for analyses of hard scattering processes at proton-(anti-)proton colliders like the TEVATRON and the LHC. Structure functions are also among the quantities best suited for precisely measuring the strong coupling constant ${\alpha}_S$. The SLAC-MIT collaboration [1-3] observed the scaling behaviour [4] of the proton structure function in DIS. This observation established the quark-parton model (QPM) as a valid framework for the interpretation of DIS data and these DIS processes can be expressed in terms of universal parton densities. In QCD, structure functions are defined as convolution of the universal parton momentum distributions inside the proton and coefficient functions, which contain information about the boson-parton interaction. At large momentum transfers $Q^2$ and not too small-$x$, where $x$ is the fraction of proton momentum carried by the parton, QCD allows the perturbative calculation of the coefficient functions and predicts a logarithmic dependence (evolution) of the proton structure functions with $Q^2$ to higher orders in ${\alpha}_S$. Thus measurements of structure functions allow perturbative QCD to be precisely tested.  The standard and the basic tools for theoretical investigation of DIS structure functions are the DGLAP evolution equations [5]. 

\ The solutions of the unpolarized DGLAP equation for the QCD evolution of parton distribution functions have been discussed considerably over the past years [6-15]. There exist two main classes of approaches: those that solve the equation directly in $x$-space, and those that solve it for Mellin transforms of the parton densities and subsequently invert the transforms back to $x$-space. Some available programs that deal with DGLAP evolution are CANDIA [11] based on the logarithmic expansions in $x$-space, QCD PEGASUS [12], which is based on the use of Mellin moments, HOPPET [14] and QCDNUM [15]. QCD PEGASUS which is a parton distribution functions [PDFs] evolution program based on Mellin-space inversion, has been used by the QCD Working Group to set some benchmark results [16-17]. HOPPET is an $x$-space evolution code that is novel both in terms of the accuracy and in terms of speed that it provides compared to other $x$-space codes. Among various methods used for the solution of this equation, most of the methods are numerical. Mellin moment space with subsequent inversion, Brute-Force method, Laguerre method etc. [18-20] are different methods used to solve DGLAP evolution equations. The shortcomings common to all are the computer time required and decreasing accuracy for $x\to0$. For example, in the Brute-Force method the accuracy is better than 1\% in the region ${10^{-5}<{x}<0.8}$. However, it takes significant amount of computing time [21]. The Laguerre method has an advantage of computing time. However, the accuracy is slightly worse at small $x$ [22]. More precise approach is the matrix approach to the solution of the DGLAP evolution equations, yet it is also a numerical solution. The matrix method is a matrix-based approach to numerical integration of the DGLAP evolution equations [23]. The method arises naturally on discretisation of the Bjorken $x$ variable which is a necessary procedure for numerical integration. In Ref. [24], a quadratic contour has been derived for calculating the evolution of parton distribution functions within the Mellin transform method, and demonstrated its superiority over other techniques in the literature. With this method the parton distribution functions can be evolved according to any of the commonly used truncations of the evolution. Moreover, this method may also be used to extract parton distribution functions from collider data. Again, in Ref. [25], Stefan Weinzierl reports on a numerical program for the evolution of parton distributions which uses the Mellin-transform method with an optimized contour. An original implementation of a very elegant numerical treatment used to solve the DGLAP evolution equations at NLO is presented in Ref. [26]. This method consists in expanding parton distributions and splitting functions on Laguerre polynomials, which reduces DGLAP equations to a set of ordinary differential equations defined by recurrence. Recently, the NNPDF collaboration [27] presented and utilized an improved fast algorithm for the solution of evolution equations and the computation of general hadronic processes. As an alternative to the numerical solution, one can study the behavior of quarks and gluons via analytic solutions of the evolution equations. Although exact analytic solutions of the DGLAP equations cannot be obtained in the entire range of $x$ and $Q^2$, such solutions are possible under certain conditions [28-29].  Some approximated analytical solutions of DGLAP evolution equations suitable at small-$x$, have been reported in recent years [6-9] [30-32] with considerable phenomenological success. For example, in Ref. [6] the spin-dependent singlet and non-singlet structure functions have been obtained by solving DGLAP evolution equations at LO and NLO in the small-x limit by using Taylor series expansion and then the method of characteristics. They have calculated the $t$- and $x$-evolutions of deuteron structure function and the results show very good agreement with the SLAC E-143 Collaboration data. Similarly, in Ref. [7], the spin-independent singlet and non-singlet structure functions have been obtained by solving the DGLAP evolution equations by using Taylor expansion and then the method of characteristics at LO and NLO. Here also the computed values of the $t$- and $x$-evolutions of deuteron structure function are in good agreement with the NMC data. An approximate solution of the DGLAP equation by using Taylor expansion is presented and the $x$-distribution of the deuteron structure function is calculated at small-$x$ in Ref. [10]. Here The results are compared with the EMC NA 28 experiment data and it shows good phenomenological success. The one-loop splitting functions corresponding to LO DGLAP equation is given in Ref. [33]. Similarly the two-loop splitting functions governing the evolution have been known for a long time [34]. The calculation of the NNLO QCD approximation for the structure functions $F_2$ of DIS is important for the understanding of perturbative QCD and for an accurate comparison of perturbative QCD with experiment. To obtain the NNLO approximation for these structure functions one needs the corresponding 3-loop splitting functions. Recently the three-loop splitting functions governing the evolutions of both the non-singlet and singlet structure functions have been reported which are in with good agreement with the existing data [35-40].

\ In a recent paper [8] the non-singlet structure functions have been obtained by solving DGLAP evolution equations at NNLO in the small-$x$ limit. Here a Taylor series expansion has been used to solve the evolution equations and to obtain the semi-numerical solution. The results are compared with the Fermilab experiment E665 data and New Muon Collaboration (NMC) data. In the present paper we extend the work to the singlet case. Here we intend to solve the DGLAP equation analytically by Taylor expansion method to obtain a semi-numerical solution for singlet structure functions at small-$x$ at NNLO. In this work we have also calculated $t$- and $x$-evolutions of deuteron structure functions and the results are compared with E665 data [41] and NMC data [42]. We have also compared our results with those obtained by the fit to $F_{2}^{d}$ produced by the NNPDF collaboration  [27, 43] in 2002 based on the NMC and BCDMS data. The NNPDF parametrizations have presented a determination of the probability density in the space of structure functions for the $F_{2}$ structure function for proton, deuteron and non-singlet, as determined from experimental data of the NMC and BCDMS collaborations. Their results take the form of a set of 1000 neural nets, for each of the three structure functions, each of which gives a determination of $F_{2}$ for given $x$ and $Q^{2}$. The central value and the errors of the structure functions determined in the NNPDF fit can be computed out of the ensemble of 1000 nets according to standard Monte Carlo techniques.  

\ The spin-dependent structure functions of the nucleon, $g_1$ and $g_2$, can be measured in polarized DIS [44] by using polarized lepton beams scattered by polarized targets. The first experiment in polarized electron-polarized proton scattering was performed in the 1970s which helped to establish the parton structure of the proton. Polarized DIS lepton nucleon scattering experiments have been performed mainly at CERN, SLAC, DESY, and JLAB, and these processes have played a key role in understanding QCD and the spin structure of the nucleon [45-51]. The solutions of spin-dependent DGLAP evolution equations give polarized quark and gluon parton densities which ultimately give polarized proton and neutron structure functions. The Taylor series expansion method that gives acceptable solutions for spin-independent evolution equations [8,10], which we use in our present work, can also be used for the spin-dependent case [52]. The LO evolution equation of polarized structure functions implies scaling violations. The NLO and higher order evolution equations of polarized structure functions will help us to understand the higher twist contributions to the scaling violations and, ultimately, the spin content of the nucleon. 

\section{Solution of singlet structure function at \\ NNLO}
\paragraph\ The singlet quark density of a hadron is given by [36]
\begin{equation} q_{S}(x,{{\mu}_f}^2,{{\mu}_r}^2)=\sum_{i=1} ^{N_f}[q_i (x,{{\mu}_f}^2,{{\mu}_r}^2) + \overline{q_i}(x,{{\mu}_f}^2,{{\mu}_r}^2)],\end{equation}
where ${q_i (x,{{\mu}_f}^2,{{\mu}_r}^2)}$ and ${\overline{q_i}(x,{{\mu}_f}^2,{{\mu}_r}^2)}$ represent the number distribution of quarks and antiquarks, respectively, in the fractional hadron momentum $x$. The corresponding gluon distribution is denoted by ${g(x,{{\mu}_f}^2,{{\mu}_r}^2)}$. The subscripts $i$ indicate the flavor of the quarks (and antiquarks), and $N_f$ is the number of effectively massless flavours, and finally $\mu_r$ and $\mu_f$ represent the renormalization and factorization scales respectively. Since $\mu_r$ and $\mu_f$ are arbitrary, their values may be chosen independently.

\  The DGLAP evolution equation in the singlet sector in the standard form is given by
\begin{equation}\frac{\partial}{{\partial}\ln{{Q}^2}}  \left( \begin{array}{c}
q_{S} \\
g   
\end{array} \right)
=\left( \begin{array}{cc}
P_{qq} & P_{qg} \\
P_{gq} & P_{gg}   
\end{array} \right) \bigotimes \left( \begin{array}{c}
q_{S} \\
g   
\end{array} \right)
\end{equation}
where $\bigotimes$ represents the standard Mellin Convolution in the momentum variable and the notation is given by  
\begin{equation}a(x)\bigotimes b(x) \equiv \int_{x}^{1} \frac{d\omega}{\omega}a(\omega)b(\frac{x}{\omega})\end{equation}

\ Thus, using equation (3), equation (2) can be written as
\begin{displaymath}\frac{\partial}{{\partial}\ln{{Q}^2}}  \left( \begin{array}{c}
q_{S} \\
g   
\end{array} \right)
=\int_{x}^{1} \frac{d\omega}{\omega}\left( \begin{array}{cc}
P_{qq} & P_{qg} \\
P_{gq} & P_{gg}   
\end{array} \right)\left( \begin{array}{c}
q_{S} \\
g   
\end{array} \right)\end{displaymath}
\begin{equation}=\int_{x}^{1} \frac{d\omega}{\omega}\left( \begin{array}{cc}
P_{qq}q_S + P_{qg}g \\
P_{gq}q_S + P_{gg}g   
\end{array} \right)\end{equation}

which implies, \begin{equation}\frac{\partial{q_S}}{{\partial}\ln{{Q}^2}}=\int_{x}^{1} \frac{d\omega}{\omega}(P_{qq}q_S + P_{qg}g)\end{equation} 
\begin{equation}\frac{\partial{g}}{{\partial}\ln{{Q}^2}}=\int_{x}^{1} \frac{d\omega}{\omega}(P_{gq}q_S + P_{gg}g)\end{equation}
where, $q_S$ and $g$ are singlet and gluon parton densities respectively and $P_{qq}$, $P_{qg}$, $P_{gq}$, $P_{gg}$ are splitting functions. The quark-quark splitting function $P_{qq}$ can be expressed as [53]
\begin{displaymath}{P_{qq}(x,Q^2)=\frac{{\alpha_S}{(Q^2)}}{2\pi}P_{qq}^{(0)}(x)+\Big(\frac{{\alpha_{S}}{(Q^{2})}}{2\pi}\Big)^2P_{qq}^{(1)}(x)+\Big(\frac{{\alpha_{S}}{(Q^{2})}}{2\pi}\Big)^3P_{qq}^{(2)}(x)}\end{displaymath}
\begin{equation}{+\mathcal{O}(P_{qq}^{(3)}(x))}\end{equation}
where, $P_{qq}^{(0)}(x)$, $P_{qq}^{(1)}(x)$ and $P_{qq}^{(2)}(x)$ are LO, NLO and NNLO quark-quark splitting functions respectively. Other splitting functions can be expressed in the same way.

\ The splitting functions can be obtained from the N-space results of the Mellin space by an inverse Mellin transformation. The quark-quark splitting function $P_{qq}$ in equation (2) can be expressed as $P_{qq} = P_{NS}^{+}+ N_f(P_{qq}^S + P_{\overline{q}q}^S) = P_{NS}^{+} + P_{PS}$. Here $P_{NS}^+$  is the non-singlet splitting function calculated up to the third order [38]. $P_{qq}^S$ and $P_{\overline{q}q}^S$ are the flavor-independent contributions to the quark-quark and quark-antiquark splitting functions respectively. At very small-$x$ the pure singlet term  $P_{PS}$ dominates over $P_{NS}^+$ [40]. The gluon-quark ($P_{gq}$) and quark-gluon ($P_{qg}$) entries in equation (2) are given by $P_{qg} = N_{f}P_{q_{i}g}$ and $P_{gq} = P_{g{q_{i}}}$, where $P_{q_{i}g}$ and $P_{gq_{i}}$ are the flavor-independent splitting functions.

\ Substituting the LO splitting functions in equation (5) and simplifying, the DGLAP evolution equations for singlet structure function at LO can be written as 
\begin{equation}\frac{{\partial}F_{2}^{S}(x,t)}{{\partial}t}-\frac{{{\alpha}_S}(t)}{2{\pi}}\Big[\frac{2}{3}\{3 + 4\ln(1-x)\}F_{2}^{S}(x,t)+I_{1}^{S}(x,t)\Big]=0 ,\end{equation}

where, \begin{equation}F_{2}^{S}(x,t)\equiv{q_{S}}=\sum_{i=1} ^{N_f}[q_i+\overline{q_i}],\end{equation}
\begin{displaymath}{{I_1^{S}(x,t)}={{\frac{4}{3}}{\int_x^1{{\frac{d{\omega}}{1-{\omega}}}[(1+{\omega}^{2})F_{2}^{S}({\frac{x}{\omega}},t)-2F_{2}^{S}(x,t)]}}}}\end{displaymath}\begin{equation}{+N_f{\int_x^1\{\omega^2+(1-\omega)^2\}G\Big({\frac{x}{\omega},t}\Big)d{\omega}}}.\end{equation}
Here $t=\ln{\dfrac{Q^{2}}{{\Lambda}^{2}}}, {\Lambda}$ is the QCD cut off parameter.  It is the scale at which partons turn themselves into hadrons. Also $G\Big({\frac{x}{\omega},t}\Big)\equiv{g}$ is the gluon parton densities.

\ Similarly, substituting the splitting functions upto NLO and upto NNLO in equation (5) and simplifying, we get the DGLAP evolution equations for singlet structure function at NLO and NNLO respectively as,

\begin{displaymath}\frac{{\partial}F_{2}^{S}(x,t)}{{\partial}t}-\frac{{{\alpha}_S}(t)}{2{\pi}}\Big[\frac{2}{3}\{3 + 4\ln(1-x)\}F_{2}^{S}(x,t)+I_{1}^{S}(x,t)\Big]\end{displaymath}\begin{equation}{-\Big(\frac{{{\alpha}_S}(t)}{2{\pi}}\Big)^{2}I_{2}^{S}(x,t)=0} ,\end{equation}

\begin{displaymath}\frac{{\partial}F_{2}^{S}(x,t)}{{\partial}t}-\frac{{{\alpha}_S}(t)}{2{\pi}}\Big[\frac{2}{3}\{3 + 4\ln(1-x)\}F_{2}^{S}(x,t)+I_{1}^{S}(x,t)\Big]\end{displaymath}\begin{equation}{-\Big(\frac{{{\alpha}_S}(t)}{2{\pi}}\Big)^{2}I_{2}^{S}(x,t)-\Big(\frac{{{\alpha}_S}(t)}{2{\pi}}\Big)^{3}I_{3}^{S}(x,t)=0} ,\end{equation}

where,

\begin{displaymath}{{I_{2}^{S}(x,t)}={{(x-1){F_{2}^{S}(x,t)}{\int_{0}^{1}{f(\omega)d\omega}}+{\int_{x}^{1}{f(\omega){F_{2}^{S}{({\frac{x}{\omega}},t)}}d{\omega}}}}}}\end{displaymath}\begin{equation}{+\int_{x}^{1}F_{qq}^S(\omega)F_2^S\Big({\frac{x}{\omega},t}\Big)d{\omega}+\int_{x}^{1}F_{qg}^S(\omega)G\Big({\frac{x}{\omega},t}\Big)d{\omega}},\end{equation}
\begin{equation}{{I_{3}^{S}(x,t)}={\int_{x}^{1}\frac{d{\omega}}{\omega}\Big[P_{qq}(x)F_{2}^{S}\Big(\frac{x}{\omega},t\Big)+P_{qg}(x)G\Big(\frac{x}{\omega},t\Big)\Big]}}.\end{equation}

\ The explicit forms of the third order splitting functions are given in Appendix A. 

\ The strong coupling constant, ${\alpha_{S}}{(Q^{2})}$ is related with the ${\beta}$-function as [54]
\begin{equation}{{{\beta(\alpha_{S})}={\dfrac{\partial\alpha_{S}(Q)^{2}}{\partial log Q^{2}}}={-{{\dfrac{\beta_{0}}{4\pi}}{\alpha_{S}^{2}}}-{{\dfrac{\beta_{1}}{16\pi^{2}}}{\alpha_{S}^{3}}}-{{{\dfrac{\beta_{2}}{64\pi^{2}}}{\alpha_{S}^{4}}}+\mathcal{O}(\alpha_S^5)}}}}\end{equation}
where, 
\begin{equation}{{\beta_{0}}={{\dfrac{11}{3}}{N_{c}}}-{{\dfrac{4}{3}}{T_{f}}}={11-{{\dfrac{2}{3}}{N_{f}}}}},\end{equation}
\begin{equation}{{\beta_{1}}={{{\dfrac{34}{3}}{N_{c}^{2}}}-{{\dfrac{10}{3}}{N_{c}N_{f}}-{2C_{F}{N_{f}}}}={102-{{{\dfrac{38}{3}}{N_{f}}}}}}}\end{equation}
\begin{displaymath}{{\beta_{2}}={{{\dfrac{2857}{54}}{N_{c}^{3}}}+{2{C_{F}^{2}}{T_{f}}}-{{\dfrac{205}{9}}{{C_{F}}{N_{c}}{T_{f}}}}+{{\dfrac{44}{9}}{{C_{F}}{T_{f}^{2}}}}+{{\dfrac{158}{27}}{{N_{c}}{T_{f}^{2}}}}}}\end{displaymath}
\begin{equation}{={{{\dfrac{2857}{6}}-{{\dfrac{6673}{18}}{N_{f}}}+{{\dfrac{325}{54}}{N_{f}^{2}}}}}}\end{equation}
\vspace{5pt}
are the one-loop, two-loop and three-loop corrections to the QCD $\beta$-function and $N_f$ being the number of quark flavor. We have used $N_{c}=3$, ${{C_{F}}={\dfrac{{N_{c}^{2}}-1}{2{N_{c}}}}}$ $={\dfrac{4}{3}}$ and ${T_{f}}={{\dfrac{1}{2}}{N_{f}}}.$

\ Now let us introduce the variable $u=1-\omega$ as discussed earlier [6-8]. Therefore, we can write ${\frac{x}{\omega}={\frac{x}{(1-u)}}=x+\frac{xu}{1-u}}$. Thus we can expand ${F_2^S(x,t)}$ by Taylor expansion method as

\vspace{10pt}

\begin{displaymath}{F_{2}^{S}(\frac{x}{\omega},t)= F_{2}^{S} (x+\frac{xu}{1-u},t)}\end{displaymath}
\begin{equation}{ =F_{2}^{S}(x,t)+ \Big(\frac{xu}{1-u}\Big) \frac{\partial{F_{2}^{S}}(x,t)}{\partial x}+\frac{1}{2}{\Big(\frac{xu}{1-u}\Big)}^2\frac{\partial^2{F_{2}^{S}}(x,t)}{\partial^2 x}+\cdots}\end{equation}

\vspace{10pt}

\ As $x$ is small in our region of discussion, the terms containing $x^2$ and higher powers of $x$ can be neglected as those terms are still smaller and therefore, we can rewrite

\begin{equation}{{F_{2}^{S}({\frac{x}{\omega}},t)}={{F_{2}^{S}(x,t)}+{{\frac{xu}{1-u}}{\frac{\partial{F_{2}^{S}}(x,t)}{\partial x}}}}}.\end{equation}

Similarly,

\begin{equation}{G(\frac{x}{\omega},t)=G(x,t)+\frac{xu}{1-u}\frac{\partial{G(x,t)}}{\partial x}}.\end{equation}

\ Using equations (20) and (21) in equations (10), (13) and (14), and performing $u$-integrations, equations (8), (11) and (12) become 
\begin{displaymath}{\frac{\partial{F_{2}^{S}}(x,t)}{\partial t}-\frac{\alpha_{S}(t)}{2\pi}\Big[A_{1}(x) F_{2}^{S}(x,t)+ A_{2}(x) \frac{\partial{F_{2}^{S}}(x,t)}{ \partial{x}} +A_{3}(x)G(x,t)}\end{displaymath}\begin{equation}{+ A_{4}(x)\frac{\partial{G(x,t)}}{\partial{x}}\Big]=0,}\end{equation}

\begin{displaymath}
{\frac{\partial{F_{2}^{S}}(x,t)}{\partial t}- \frac{\alpha_{S}(t)}{2\pi}\Big[A_{1}(x) F_{2}^{S}(x,t)+ A_{2}(x) \frac{\partial{F_{2}^{S}}(x,t)}{ \partial{x}} +A_{3}(x)G(x,t)}\end{displaymath}
\begin{displaymath}{{+ A_{4}(x)\frac{\partial{G(x,t)}}{\partial{x}}\Big]}+ \Big(\frac{\alpha_{S}(t)}{2\pi}\Big)^2\Big[B_{1}(x) F_{2}^{S}(x,t)}\end{displaymath}
\begin{equation}{+ B_{2}(x) \frac{\partial{F_{2}^{S}}(x,t)}{ \partial{x}}+ B_{4}(x)\frac{\partial{G(x,t)}}{\partial{x}}\Big]=0}
\end{equation}

\begin{eqnarray}\nonumber
{\frac{\partial{F_{2}^{S}}(x,t)}{\partial t}- \frac{\alpha_{S}(t)}{2\pi}\Big[A_{1}(x) F_{2}^{S}(x,t)+ A_{2}(x)\frac{\partial{F_{2}^{S}}(x,t)}{ \partial{x}}+A_{3}(x)G(x,t)}\\\nonumber
{+ A_{4}(x)\frac{\partial{G(x,t)}}{\partial{x}}\Big]+ \Big(\frac{\alpha_{S}(t)}{2\pi}\Big)^2\Big[B_{1}(x) F_{2}^{S}(x,t)+ B_{2}(x) \frac{\partial{F_{2}^{S}}(x,t)}{ \partial{x}}}\\\nonumber
{+B_{3}(x)G(x,t)+ B_{4}(x)\frac{\partial{G(x,t)}}{\partial{x}}\Big]+\Big(\frac{\alpha_{S}(t)}{2\pi}\Big)^3\Big[C_{1}(x) F_{2}^{S}(x,t)}\\
{+ C_{2}(x) \frac{\partial{F_{2}^{S}}(x,t)}{ \partial{x}}+C_{3}(x)G(x,t)+ C_{4}(x)\frac{\partial{G(x,t)}}{\partial{x}}\Big]=0}
\end{eqnarray}

\ where, $A_i(x)$, $B_i(x)$ and $C_i(x)$ (where $i$=1,2,3,4) are functions of $x$ (see Appendix B).

\ The $Q^2$-evolution of the proton structure function $F_2(x,Q^2)$ is related to the gluon parton densities in the proton, $G(x,Q^2)$, and to the strong interaction coupling constant, $\alpha_S$. The gluon parton densities cannot be measured directly through experiments. It is, therefore, important to measure the gluon parton densities $G(x,Q^2)$ indirectly using the proton structure functions $F_2(x,Q^2)$. Hence the direct relations between $F_2(x,Q^2)$ and the gluon parton densities $G(x,Q^2)$ are extremely important; using those relations the experimental values of $G(x,Q^2)$ can be extracted using the data on $F_2(x,Q^2)$. Therefore, in the analytical solutions of DGLAP evolution equations for singlet structure functions or gluon parton densities, a relation between singlet structure function and gluon parton densities has to be assumed [6,7]. The commonly used relation is ${G(x,t)=K(x)F_2^S(x,t)}$, where $K(x)$ is an ad hoc function of $x$. Since these evolution equations of gluon parton densities and singlet structure functions are in the same forms of derivative with respect to $t$, so we can consider this function. And also the input singlet and gluon parameterizations, taken from global analysis of parton distribution functions, in particular from the MSTW08 parton set, to incorporate different high precision data, are also functions of $x$ at fixed $Q^2$ [55]. So the relation between singlet structure function and gluon parton densities will come out in terms of $x$ at fixed-$Q^2$. However the actual functional form of $K(x)$ can be determined by simultaneous solutions of coupled equations of singlet structure functions and gluon parton densities.

\  Hence equations (22), (23) and (24) take the form

\begin{equation}{-t\frac{\partial{F_2^S(x,t)}}{\partial{t}}+L_1(x,t) \frac{\partial{F_2^S(x,t)}}{\partial{x}}+M_1(x,t)F_2^S(x,t)=0},\end{equation}

\begin{equation}{-t\frac{\partial{F_2^S(x,t)}}{\partial{t}}+L_2(x,t) \frac{\partial{F_2^S(x,t)}}{\partial{x}}+M_2(x,t)F_2^S(x,t)=0},\end{equation}

\begin{equation}{-t\frac{\partial{F_2^S(x,t)}}{\partial{t}}+L_3(x,t) \frac{\partial{F_2^S(x,t)}}{\partial{x}}+M_3(x,t)F_2^S(x,t)=0},\end{equation}
where 
\begin{equation}{L_1(x,t)=\frac{3}{2}A_f(t)[A_2(x)+K(x)A_4(x)]},\end{equation} 
\begin{equation}{M_1(x,t)=\frac{3}{2}A_f(t)\big[A_1(x)+K(x)A_3(x)+\frac{\partial{K(x)}}{\partial{x}}A_4(x)\big]},\end{equation}
\begin{equation}{L_2(x,t)=\frac{3}{2}A_f(t)[(A_2(x)+K(x)A_4(x))+T_0(B_2(x)+K(x)B_4(x))]},\end{equation}
\begin{displaymath}{M_2(x,t)=\frac{3}{2}A_f(t)[\big(A_1(x)+K(x)A_3(x)+\frac{\partial{K(x)}}{\partial{x}}A_4(x)\big)+T_0 \big(B_1(x)}\end{displaymath}
\begin{equation}{+K(x)B_3(x)+\frac{\partial{K(x)}}{\partial{x}}B_4(x)\big)]},\end{equation}
\begin{displaymath}{L_3(x,t)=\frac{3}{2}A_f(t)[(A_2(x)+K(x)A_4(x))+T_0(B_2(x)+K(x)B_4(x))}\end{displaymath}
\begin{equation}{+T_1(C_2(x)+K(x)C_4(x))]},\end{equation}
\begin{displaymath}{M_3(x,t)=\frac{3}{2}A_f(t)[\big(A_1(x)+K(x)A_3(x)+\frac{\partial{K(x)}}{\partial{x}}A_4(x)\big)+T_0 \big(B_1(x)}\end{displaymath}
\begin{displaymath}{+K(x)B_3(x)+\frac{\partial{K(x)}}{\partial{x}}B_4(x)\big)+ T_1\big(C_1(x)+K(x)C_3(x)}\end{displaymath}
\begin{equation}{+\frac{\partial{K(x)}}{\partial{x}}C_4(x)\big)]}\end{equation}
and
\begin{equation}{\frac{\alpha_S(t)}{2\pi}= \frac{3A_f(t)}{2t}}.\end{equation}

\ Also one can consider two numerical parameters $T_0$ and $T_1$, such that  ${T^2(t)=T_0T(t)}$ and ${T^3(t)=T_1T(t)},$ where ${T(t)= \frac{\alpha_S(t)}{2\pi}}$. As discussed in [8], here also we have considered $T_0= 0.05$ and $T_1= 0.006$ within the range ${0\leq{Q^2}\leq30}$ GeV$^2$. These values are chosen such that differences  between $T^2(t)$  and $T_0T(t)$, and also $T^3(t)$ and $T_1T(t)$ are  negligible. Thus the consideration of the parameters $T_0$ and $T_1$ does not give any abrupt change in our result.     

\ The general solution of the equations (25), (26) and (27) is ${F(U,V) = 0},$ where ${F(U,V)}$ is an arbitrary function. Here, ${U(x, t, F_2^S) = K_1}$ and ${V(x, t, F_2^S)}$ \\ ${=K_2}$ are two independent solutions of the equation
\begin{equation}{\frac{\partial{x}}{L_i(x,t)}=\frac{\partial{t}}{-t}=\frac{\partial{F_2^S(x,t)}}{-M_i(x,t)F_2^S(x,t)}}.\end{equation}

\ To get the solution of equations (25), (26) and (27) we will follow the same procedure as discussed in Ref. [8]. Now, let us introduce functions $\overline{L_i}(x)$ and $\overline{M_i}(x)$ such that, ${L_i(x,t)=\frac{{\beta_0}t}{2}T(t)\overline{L_i}(x)}$ and ${M_i(x,t)=\frac{{\beta_0}t}{2}T(t)\overline{M_i}(x)}$  where i=1,2,3 for LO, NLO and NNLO respectively. 

\ Solving this equation for LO, we obtain 
\begin{equation}{ U(x, t, F_2^S) = t\cdot{\exp}\Big[\int\frac{1}{\overline{L_1}(x)}dx\Big]}\end{equation} and
\begin{equation}{ V(x, t, F_2^S) = F_2^S(x,t)\cdot{\exp}\Big[\int\frac{\overline{M_1}(x)}{\overline{L_1}(x)}dx\Big]},\end{equation}
where  \begin{equation}{\overline{L_1}(x)=\frac{2}{\beta_0}(A_2+KA_4)},\end{equation}
\begin{equation}{\overline{M_1}(x,t)=\frac{2}{\beta_0}\big(A_1+KA_3+\frac{\partial{K}}{\partial{x}}A_4\big)}.\end{equation}

\ Thus it has no unique solution. The simplest possibility to get a solution of the equation (25) is ${\alpha\cdot{U}+ \beta\cdot{V}=0},$ where $\alpha$ and $\beta$ are arbitrary constants. Putting the values of U and V in this equation we get
\begin{equation}{\alpha{t}\cdot{\exp}\Big[\int\frac{1}{\overline{L_1}(x)}dx\Big]+\beta{F_2^S(x,t)}\cdot{\exp}\Big[\int\frac{\overline{M_1}(x)}{\overline{L_1}(x)}dx\Big]=0},\end{equation}
from which we get,  \begin{equation} {F_2^S(x,t)=-\gamma{t}\cdot{\exp}\Big[\int\big(\frac{1}{\overline{L_1}(x)}- \frac{\overline{M_1}(x)}{\overline{L_1}(x)}\big)dx\Big]},\end{equation}
where ${\gamma=\frac{\alpha}{\beta}}$ is another constant.
Now defining \begin{equation}{F_2^S(x,t_0)=-\gamma{t_0}\cdot{\exp}\Big[\int\big(\frac{1}{\overline{L_1}(x)}- \frac{\overline{M_1}(x)}{\overline{L_1}(x)}\big)dx\Big],}\end{equation}
at ${t=t_0},$ where  ${t_0=\ln{\dfrac{Q^{2}}{{\Lambda}^{2}}}},$ for any lower value ${Q^2=Q_0^2},$ we get from equation (41)
 \begin{equation} {F_2^S(x,t)=F_2^S(x,t_0)\big(\frac{t}{t_0}\big)}\end{equation}
Again defining at ${x=x_0},$ \begin{equation}{F_2^S(x_0,t)=-\gamma{t}\cdot{\exp}\Big[\int\big(\frac{1}{\overline{L_1}(x)}- \frac{\overline{M_1}(x)}{\overline{L_1}(x)}\big)dx\Big]_{x=x_0}},\end{equation}
we obtain  \begin{equation} {F_2^S(x,t)=F_2^S(x_0,t)\cdot{\exp}\Big[\int_{x_0}^{x}\big(\frac{1}{\overline{L_1}(x)}- \frac{\overline{M_1}(x)}{\overline{L_1}(x)}\big)dx\Big],}\end{equation}

\ Equations (43) and (45) give the $t$- and $x$- evolutions of singlet structure functions at LO, respectively.

\ The deuteron structure function can be written in terms of the singlet quark distribution function as [3],
\begin{equation} {F_2^d(x,t)=\frac{5}{9}F_2^S(x,t)}\end{equation}

\ Substituting equations (43) and (45) in equation (46), the $t$- and $x$-evolutions of deuteron structure function at LO can be obtained as
\begin{equation} {F_2^d(x,t)=F_2^d(x,t_0)\big(\frac{t}{t_0}\big)}\end{equation}
and
\begin{equation} {F_2^d(x,t)=F_2^d(x_0,t)\cdot{\exp}\Big[\int_{x_0}^{x}\big(\frac{1}{\overline{L_1}(x)}- \frac{\overline{M_1}(x)}{\overline{L_1}(x)}\big)dx\Big],}\end{equation}

\ Proceeding in the same way we obtain the $t$ and $x$-evolution of deuteron structure function ${F_2^d(x,t)}$ at NLO as
\begin{equation} {F_2^d(x,t)=F_2^d(x,t_0)\Big(\frac{t^{1+{b/t}}}{t_0^{1+{b/t_0}}}\Big)\cdot{\exp}\Big(\frac{b}{t}-\frac{b}{t_0}\Big)}\end{equation}
and 
\begin{equation} {F_2^d(x,t)=F_2^d(x_0,t)\cdot{\exp}\Big[\int_{x_0}^{x}\big(\frac{1}{\overline{L_2}(x)}- \frac{\overline{M_2}(x)}{\overline{L_2}(x)}\big)dx\Big],}\end{equation}
where
\begin{equation}{F_2^d(x,t_0)=\frac{5}{9}\Big\{-\gamma{t_0^{(1+b/t_0)}}\cdot{\exp}(\frac{b}{t_0})\cdot{\exp}\Big[\int\big(\frac{1}{\overline{L_2}(x)}- \frac{\overline{M_2}(x)}{\overline{L_2}(x)}\big)dx\Big]\Big\}},\end{equation}
\begin{displaymath}{F_2^d(x_0,t)=\frac{5}{9}\Big\{-\gamma{t^{(1+b/t)}}\cdot{\exp}\Big(\frac{b}{t}\Big)\cdot{\exp}\Big[\int\big(\frac{1}{\overline{L_2}(x)}}\end{displaymath}
\begin{equation}{-\frac{\overline{M_2}(x)}{\overline{L_2}(x)}\big)dx\Big]_{x=x_0}\Big\}},\end{equation}
\begin{equation}{\overline{L_2}(x,t)=\frac{2}{\beta_0}[(A_2+KA_4)+T_0(B_2+KB_4)]},\end{equation}
\begin{equation}{\overline{M_2}(x,t)=\frac{2}{\beta_0}[\big(A_1+KA_3+\frac{\partial{K}}{\partial{x}}A_4\big)+T_0 \big(B_1+KB_3+\frac{\partial{K}}{\partial{x}}B_4\big)]},\end{equation}

\ Similarly, the $t$- and $x$-evolutions of deuteron structure function at NNLO can be obtained as
\begin{displaymath} {F_2^d(x,t)=F_2^d(x,t_0)\Big(\frac{t^{1+{(b-b^2)/t}}}{t_0^{1+{(b-b^2)/t_0}}}\Big)\cdot{\exp}\Big(\frac{b-c-b^2\ln^2{t}}{t}}\end{displaymath}\begin{equation}{-\frac{b-c-b^2\ln^2{t_0}}{t_0}\Big)}\end{equation}
and
\begin{equation} {F_2^d(x,t)=F_2^d(x_0,t)\cdot{\exp}\Big[\int_{x_0}^{x}\big(\frac{1}{\overline{L_3}(x)}- \frac{\overline{M_3}(x)}{\overline{L_3}(x)}\big)dx\Big],}\end{equation}
where

\ \begin{displaymath}{F_2^d(x,t_0)=\frac{5}{9}\Big\{-\gamma{t_0^{(1+(b-b^2)/t_0)}}\cdot{\exp}\Big(\frac{b-c-b^2\ln^2t_0}{t_0}\Big)\cdot{\exp}\Big[\int\big(\frac{1}{\overline{L_3}(x)}}\end{displaymath}
\begin{equation}{-\frac{\overline{M_3}(x)}{\overline{L_3}(x)}\big)dx\Big]\Big\}},\end{equation}
\begin{displaymath}{F_2^d(x_0,t)=\frac{5}{9}\Big\{-\gamma{t^{(1+(b-b^2)/t)}}\cdot{\exp}\Big(\frac{b-c-b^2\ln^2t}{t}\Big)\cdot{\exp}\Big[\int\big(\frac{1}{\overline{L_3}(x)}}\end{displaymath}
\begin{equation}{-\frac{\overline{M_3}(x)}{\overline{L_3}(x)}\big)dx\Big]_{x=x_0}\Big\}},\end{equation}
\begin{equation}{\overline{L_3}(x,t)=\frac{2}{\beta_0}[(A_2+KA_4)+T_0(B_2+KB_4) +T_1(C_2+KC_4)]},\end{equation}
\begin{displaymath}{\overline{M_3}(x,t)=\frac{2}{\beta_0}[\big(A_1+KA_3+\frac{\partial{K}}{\partial{x}}A_4\big)+T_0 \big(B_1+KB_3+\frac{\partial{K}}{\partial{x}}B_4\big)}\end{displaymath}
\begin{equation}{+ T_1\big(C_1+KC_3+\frac{\partial{K}}{\partial{x}}C_4\big)]},\end{equation}

\ with ${b=\frac{\beta_1}{\beta_0^2}\ {\rm and} \ c=\frac{\beta_2}{\beta_0^3}}\cdot$

\ Here the input for $F_2^d(x,t_0)$ is taken from experimental data corresponding to the lowest-$Q^2$ point. Similarly, the input for  $F_2^d(x_0,t)$ is taken from the experimental data corresponding to the highest-$x$ point.

\ HERA shows that the structure function has a steep behavior in the small-x region  ${10^{-5}<{x}<10^{-2}}$. This steep behavior is well described in the framework of the DGLAP evolution equations. So we can expect that the method we used in this work to solve the DGLAP evolution equations is valid in the small-x region; roughly in the region ${10^{-3}<{x}<10^{-1}}$. This method may also be applied in the region where ${x<10^{-3}}$, but we have not checked the validity of the method in this region. However, our method is not valid at very small-x where recombination processes between gluons have to be taken into account, since DGLAP equations fail to describe the recombination processes. 

\section{Result and discussion}
\paragraph \ In this paper we have calculated the $t$- and $x$-evolutions of singlet structure functions. The deuteron structure function is directly related to the singlet structure function as given by the equation (46). To determine the proton structure function we need to know both the singlet and non-singlet structure functions. Therefore in this particular work we have compared the results only to the deuteron structure function measured in fixed-target experiments. We have compared our results of $t$- and $x$-evolutions of deuteron structure function  with NMC and E665 experimental data in figure 1(a,b) and figure 2(a,b) respectively. In figure 3(a,b) we have compared our results with thosed obtained by NNPDF parameterizations. We consider the range ${0.0052\leq{x}\leq0.18}$ and ${1.094\leq{Q^2}\leq26}$ GeV$^2$ for E665 data, ${0.0045\leq{x}}$ ${\leq0.18}$ and ${0.75\leq{Q^2}\leq27}$ GeV$^2$ for NMC data. We have used the range ${0.0045\leq{x}}$ ${\leq0.09}$ and $1.25\leq{Q^2}\leq26$ GeV$^2$ to compare our results with those obtained by the fit to $F_{2}^{d}$ produced by the NNPDF collaboration in 2002 based on the NMC and BCDMS data. In the $t$-evolutions of deuteron  structure  function  our  computed  values  of $F_2^d(x,t)$ from equations (47), (49) and (55) for LO, NLO and NNLO respectively are plotted against $Q^2$ for  different values of $x$. On the other hand, for $x$-evolutions our  computed values of $F_2^d(x,t)$ from equations (48), (50) and (56) for LO, NLO and NNLO respectively are  plotted   against  $x$  for different  values  of   $Q^2$.  For convenience, value of each data point is increased by adding 0.5$i$ or 0.4$i$ or 0.3$i$ or 0.2$i$ or 0.1$i$ where $i = 0, 1, 2,3,\cdots$ are the numberings of curves counting from the bottom of the lowermost curve as the 0-th order. In all graphs, the lowest-$Q^2$ and highest-$x$ points are taken as inputs for $F_2^d(x,t_0)$ and $F_2^d(x_0,t)$ respectively. Here vertical error bars are total statistical and systematic uncertainties. It is observed that the best-fit curves are obtained for ${2\leq{K}\leq8}$. Here we have also performed a chi-square test of our computed results and we got chi-square per degree of freedom of about 0.9.

\ We observe that the $t$- and $x$-evolutions of deuteron structure function are in good consistency with the  experimental data and parametrizations. As observed from this work it is clear that the region of validity of our method is approximately in the range ${10^{-3}\leq{x}\leq10^{-1}}$ and $0.5\leq{Q^2}\leq30$ GeV$^2$. But this methed is also valid for other ranges of $x$ and $Q^{2}$. Though various methods like Laguerre polynomials, Brute-Force method, Mellin transformation etc. are available in order to obtain a numerical solution of DGLAP evolution equations, our method to solve these equations analytically is also a workable alternative. Here we consider a parameter  $K(x)$ in assuming a relation between singlet structiure function and gluon parton densities. We have also used two other parameters like $T_{0}$ and $T_{1}$. However the number of parameters used here is smaller compared to other methods where several parameters have been used mainly in input functions [55, 56]. Moreover with this method we can calculate the $x$-evolutions of deuteron structure function in addition to the $t$-evolutions.
\vspace{10pt}
\section{\bf{Acknowledgement}}

Two of the authors (M. D. and J. K. S.) are grateful to UGC for financial support in the form of a major research project (UGC's\ Letter No. F. 37-369/2009 (SR), dated 12 January, 2010). We also thank S. Ghosh for his help in typing of the manuscript.

\ 

{\bf{\large{Appendix:A}}}

\

\ In Ref. [40] the three loop splitting functions have been presented in both Mellin-N and Bjorken-x space. The exact expressions given in equations  (4.12) and (4.13) in reference [40]  for the functions ${P_{PS}^{(2)}}(x)$ and ${P_{qg}^{(2)}}(x)$ are not simple to use. Therefore they have also presented compact approximate representations. As it is shown in reference [40], at small-x ($x\to0$) the coefficients of the $1/x$ terms in the definition of ${P_{PS}^{(2)}}(x)$ and ${P_{qg}^{(2)}}(x)$ can be written in terms of the color factors and the Riemann $\zeta$-functions. After inserting the values of the color factors and the approximated numerical values of $\zeta$-functions, the three loop splitting functions ${P_{PS}^{(2)}}(x)$ and ${P_{qg}^{(2)}}(x)$ are approximated in the limit ($x\to0$) as given in Ref. [40]. The third-order pure-singlet contribution to the quark-quark splitting function is 
\begin{displaymath}{{P_{PS}^{(2)}}(x)\cong\Big[N_f(-5.92L_1^3-9.751L_1^2-72.11L_1+177.4+392.9x-101.4x^2}\end{displaymath}
\begin{displaymath}{-57.04L_0L_1-661.61L_0+131.4L_0^2-\frac{400}{9}L_0^3+\frac{160}{27}L_0^4-506.0x^{-1}}\end{displaymath}
\begin{displaymath}{-\frac{3584}{27}x^{-1}L_0)+N_f^2(1.778L_1^2+5.944L_1+100.1-125.2x+49.26x^2-12.59x^3}\end{displaymath}
\begin{equation}{-1.889L_0L_1+61.75L_0+17.89L_0^2+\frac{32}{27}L_0^3+\frac{256}{81}x^{-1})\Big](1-x)},\end{equation}

\ The three-loop quark-gluon splitting function is
\begin{displaymath}{P_{qg}^{(2)}(x)\cong{N_f\big(\frac{100}{27}L_1^4-\frac{70}{9}L_1^3-120.5L_1^2+104.42L_1+2522-3316x}+2126x^2}\end{displaymath}
\begin{displaymath}{+L_0L_1(1823-25.22L_0)-252.5xL_0^3+424.9L_0+881.5L_0^2-\frac{44}{3}L_0^3}\end{displaymath}
\begin{displaymath}{+\frac{536}{27}L_0^4-1268.3x^{-1}-\frac{896}{3}x^{-1}L_0\big)+N_f^2\big(\frac{20}{27}L_1^3+\frac{200}{27}L_1^2-5.496L_1}\end{displaymath}
\begin{displaymath}{-252.0+158.0x+145.4x^2-98.07xL_0^2+11.70xL_0^3-L_0L_1(53.09}\end{displaymath}
\begin{equation}{+80.616L_0)-254.0L_0-90.80L_0^2-\frac{376}{27}L_0^3-\frac{16}{9}L_0^4+\frac{1112}{243}x^{-1}\big)},\end{equation}

\ The non-singlet splitting function calculated upto third order is given by 
\begin{displaymath}{{P_{NS}^{(2)}}(x)=N_f\Big[\{L_1(-163.9{x^{-1}}-7.208x)+151.49+44.51x- 43.12x^2}\end{displaymath}
\begin{displaymath}{+4.82x^3\}(1-x)+L_0L_1(-173.1+46.18L_0)+178.04L_0}\end{displaymath}
\begin{equation}{+6.892{L_0^2}+\frac{40}{27}(L_0^4-2L_0^3)}\Big],\end{equation}

\ Again, the functions ${F_{qq}^S(\omega)}$ and ${F_{qg}^S(\omega)}$ are defined as
\begin{equation}{F_{qq}^S(\omega)=2C_FT_RN_fF_{qq}(\omega)},\end{equation}
\begin{equation}{F_{qg}^S(\omega)=C_FT_RN_fF_{qg}^1(\omega)+C_GT_RN_fF_{qg}^2(\omega)},\end{equation}
where,

\begin{displaymath}{F_{qq}(\omega)=\frac{20}{9{\omega}}-2+6{\omega}-\frac{56}{9}{\omega}^2+\Big(1+5{\omega}+\frac{8}{3}{\omega}^2\Big)\ln({\omega})}\end{displaymath}
\begin{equation}{-(1+\omega)\ln^2({\omega})},\end{equation}
\begin{displaymath}{F_{qg}^1(\omega)=4-9{\omega}-(1-4{\omega})\ln(\omega)-(1-2{\omega})\ln^2(\omega)+4\ln(1-{\omega})}\end{displaymath}
\begin{equation}{+\Big[2\ln^2(\frac{1-\omega}{\omega})-4\ln(\frac{1-\omega}{\omega})-\frac{2}{3}{\pi}^2+10\Big]P_{qg}^1(\omega)},\end{equation}
\begin{displaymath}{F_{qg}^2(\omega)}={\frac{182}{9}+\frac{14}{9}{\omega}+\frac{40}{9{\omega}}+(\frac{136}{3}{\omega}-\frac{38}{3})\ln(\omega)-4\ln(1-{\omega})}\end{displaymath}
\begin{displaymath}{-(2+8{\omega})\ln^2(\omega)+\Big[-\ln^2(\omega)+\frac{44}{3}\ln(\omega)-2\ln^2(1-{\omega})}\end{displaymath}
\begin{equation}{+4\ln(1-{\omega})+\frac{\pi^2}{3}-\frac{218}{3}\Big]P_{qg}^1(\omega)+2P_{qg}^1(-{\omega})\int_{\frac{\omega}{1+{\omega}}}^{\frac{1}{1+{\omega}}}\frac{dz}{z}\ln{\frac{1-z}{z}}},\end{equation}
where, the two-loop quark-gluon splitting function $P_{qg}^{(1)}$ is defined in Ref. [34]. Here, ${T_R=\frac{1}{2}}$, ${C_G\equiv{N_C=3}}$ and ${{C_{F}}={\dfrac{{N_{c}^{2}}-1}{2{N_{c}}}}}$ $={\dfrac{4}{3}}$. 

\ Also, the function ${f(\omega)}$ is given as
\begin{displaymath}{{f(\omega)}={{{C_{F}^{2}}{[{P_{F}(\omega)}-{P_{A}(\omega)}]}+{{\dfrac{1}{2}}{{C_{F}}{C_{A}}{[{P_{G}}+{P_{A}(\omega)}]}}}+}}}\end{displaymath}
\begin{equation}{{{C_{F}}{T_{R}}{N_{f}}{P_{N_{f}}(\omega)}}},\end{equation}
where,
\begin{equation}{{P_{N_{f}}(\omega)}={{\dfrac{2}{3}}{\Big[{{\dfrac{1+{\omega}^{2}}{1-{\omega}}}{(-\ln{\omega}-{\dfrac{5}{3}})}}-{2(1-\omega)}\Big]}}},\end{equation}
\begin{displaymath}{{P_{F}(\omega)}={{-{\dfrac{2(1+{\omega}^{2})}{(1-{\omega})}{\ln(\omega)\ln(1-{\omega})}}-{\Big({\dfrac{3}{1-{\omega}}}+2{\omega}\Big)\ln{\omega}}}}}\end{displaymath}
\begin{equation}{{-{{\dfrac{1}{2}}(1+{\omega})\ln{\omega}}}+{{\dfrac{40}{3}(1-{\omega})}}},\end{equation}
\begin{displaymath}{{P_{G}(\omega)}={{\dfrac{(1+{\omega}^{2})}{(1-{\omega})}}{\Big(\ln^{2}(\omega)+{\dfrac{11}{3}}\ln(\omega)+{\dfrac{67}{9}}-{\dfrac{\pi^2}{3}}\Big)}-{\frac{1}{2}(1+{\omega})\ln{\omega}}}}\end{displaymath}
\begin{equation}{+{{\dfrac{40}{3}}(1-{\omega})}},\end{equation}
\begin{displaymath}{{P_{A}(\omega)}={{{\dfrac{2{(1+{\omega}^{2})}}{(1+{\omega})}}{{\int_{(\frac{\omega}{1+{\omega}})}^{({\frac{1}{1+{\omega}}})}{\dfrac{dk}{k}}{\ln{\Big({\dfrac{1-k}{k}}\Big)}}}}}+{2(1+{\omega})\ln({\omega})}}}\end{displaymath}
\begin{equation}{+{4(1-{\omega})}}\end{equation}

\ with $L_0=\ln(x),$  $L_1=\ln(1-x).$ Here results are from direct $x$-space evolution and  are calculated using Fortran package [35-40]. 

\ 

{\bf{\large{Appendix:B}}}

\ 

\ The explicit forms of the functions  $A_i(x)$, $B_i(x)$ and $C_i(x)$ (where $i$=1,2,3,4) are 

\begin{equation}{{A_{1}(x)}={2x+{x^{2}}+4\ln(1-x)}},\end{equation}
\begin{equation}{{A_{2}(x)}={x-{x^{3}}-2x\ln(x)}},\end{equation}
\begin{equation}{A_3(x)=2N_f\big(\frac{2}{3}-x+x^2-\frac{2}{3}x^3\big)},\end{equation}
\begin{equation}{A_4(x)=2N_f\big(-\frac{5}{3}x+3x^2-2x^3+\frac{2}{3}x^4-x\ln(x)\big)},\end{equation}
\begin{equation}{{B_{1}(x)}={x{\int_{0}^{1}{f(\omega)d\omega}}-{\int_{0}^{x}{f(\omega)d\omega}}+{{\frac{4}{3}}{N_{f}}}{\int_{x}^{1}{F_{qq}(\omega)d\omega}}}},\end{equation}
\begin{equation}{B_{2}(x)={x{\int_{x}^{1}{\big[f(\omega)}+{{\frac{4}{3}}{N_{f}}{F_{qg}^{s}(\omega)}\big]}}\frac{1-\omega}{\omega}d\omega}},\end{equation}
\begin{equation}{B_3(x)= \int_{x}^{1}F_{qg}^S(\omega)d\omega},\end{equation}
\begin{equation}{B_4(x)= x\int_{x}^{1}\frac{1-\omega}{\omega} F_{qg}^S(\omega)d\omega},\end{equation}
\begin{equation}{ C_1(x)= N_f \int_{0}^{1-x}\frac{{\omega}d{\omega}}{1-\omega}R_1(\omega)},\end{equation}
\begin{equation}{ C_2(x)= N_f \int_{0}^{1-x}\frac{{\omega}x{d\omega}}{(1-\omega)^2}R_1(\omega)},\end{equation}
\begin{equation}{ C_3(x)= N_f \int_{0}^{1-x}\frac{{\omega}x{d\omega}}{(1-\omega)^2}R_1(\omega)},\end{equation}
\begin{equation}{ C_4(x)= N_f \int_{0}^{1-x}\frac{{\omega}x{d\omega}}{(1-\omega)^2}R_2(\omega)},\end{equation}
where,
\begin{eqnarray}\nonumber 
R_1({\omega}) = & \{\ln(\omega)\ln(1-\omega)[-173.1+46.18 \ln(1-\omega)]+178.04 \ln(1-\omega)\\\nonumber
& {+6.892 \ln^2(1-\omega)+\frac{40}{27} [\ln^4(1-\omega)-2 \ln^3(1-\omega)]\}+\omega\{ \ln(\omega)}\\\nonumber
& {(-163.9(1-\omega)^{-1}-7.208(1-\omega)\big)+151.49+44.51(1-\omega)}\\\nonumber
& {-43.12(1-\omega)^2+4.82(1-\omega)^3\}+\omega^2\{-5.926 \ln^3(\omega)}\\\nonumber
& {-9.751 \ln^2(\omega)-72.11 \ln(\omega)+177.4+392.9(1-\omega)}\\\nonumber 
& {-101.4(1-\omega)^2-57.04\ln(1-\omega)\ln(\omega)-661.6 \ln(1-\omega)}\\\nonumber
& {+131.4 \ln^2(1-\omega)-\frac{400}{9} \ln^3(1-\omega)+\frac{160}{27}\ln^4(1-\omega)}\\\nonumber 
& {-506.0(1-\omega)^{-1}-\frac{3584}{27}(1-\omega)^{-1}\ln(1-\omega)\}+N_f\omega\{1.778\ln^2(\omega)}\\\nonumber
& {+5.944\ln(\omega)+100.1-125.2(1-\omega)+49.26(1-\omega)^2}\\\nonumber 
& {-12.59(1-\omega)^3-1.889\ln(1-\omega)\ln(\omega)+61.75\ln(1-\omega)} \\\
& {+17.89\ln^2(1-\omega)+\frac{32}{27}\ln^3(1-\omega)+\frac{256}{81}(1-\omega)^{-1}\}}
\end{eqnarray}
\begin{eqnarray}\nonumber
R_2(\omega)=&\{\frac{100}{27} \ln^4(\omega)-\frac{70}{9}\ln^3(\omega)-120.5\ln^2(\omega)+104.42\ln(\omega)+2522\\\nonumber
& {-3316(1-\omega)+2126(1-\omega)^2-252.5(1-\omega)\ln^3(1-\omega)}\\\nonumber
& {+ \ln(\omega)\ln(1-\omega)\big(1823-25.22\ln(1-\omega)\big)+424.9\ln(1-\omega)}\\\nonumber
& {+881.5 \ln^2(1-\omega)-\frac{44}{3} \ln^3(1-\omega)+\frac{536}{27} \ln^4(1-\omega)-1268.3}\\\nonumber
& {(1-\omega)^{-1}-\frac{896}{3}(1-\omega)^{-1}\ln(1-\omega)\}+N_f\{\frac{20}{27} \ln^3(\omega)+\frac{200}{27} \ln^2(\omega)}\\\nonumber
& {-5.496 \ln(\omega)-252.0+158.0(1-\omega)+145.4(1-\omega)^2}\\\nonumber
& {-139.28(1-\omega)^3-98.07(1-\omega) \ln^2(1-\omega)+11.70(1-\omega)}\\\nonumber
& {\ln^3(1-\omega)-\ln(\omega)\ln(1-\omega)(53.09+80.616 \ln(1-\omega))}\\\nonumber
& {-254.0\ln(1-\omega)-90.80 \ln^2(1-\omega)-\frac{376}{27} \ln^3(1-\omega)}\\\
& {-\frac{16}{9}\ln^4(1-\omega)+\frac{1112}{243}(1-\omega)^{-1}\}}
\end{eqnarray}

\begin{figure*}
   \centerline{
 \mbox{{\includegraphics[width=3.50in]{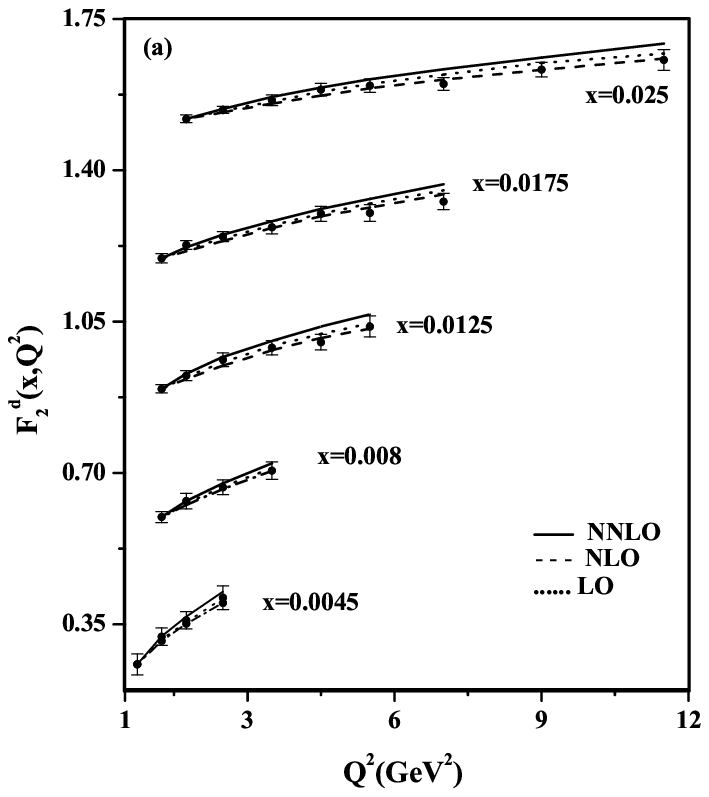}}
{\includegraphics[width=3.50in]{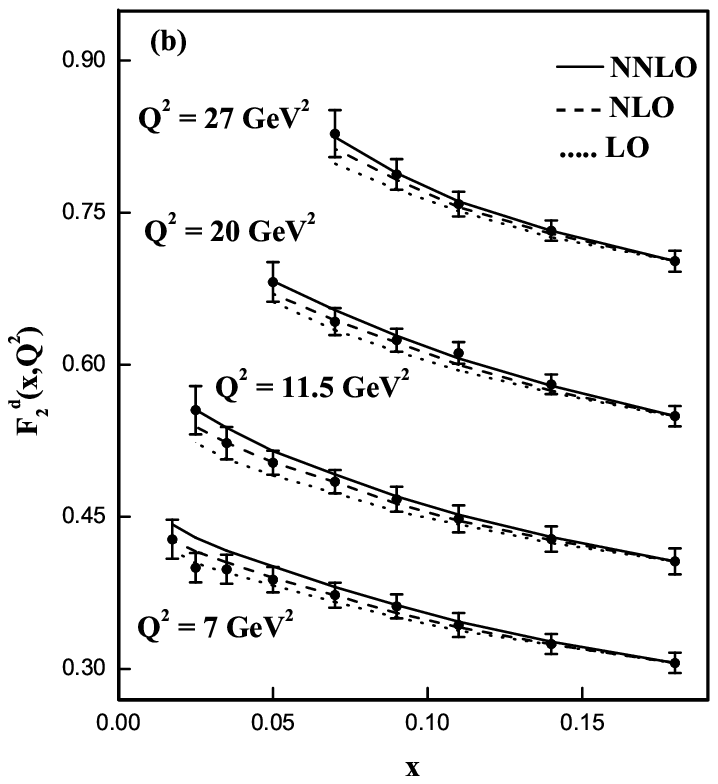}}}
  }
  \caption{NMC data. Dotted lines are LO results, dashed lines are NLO and solid lines are our NNLO results. For clarity, data are scaled up by +0.2i (in Fig.(a)) and +i (in Fig.(b)) (with i = 0,1,2,3) starting from the bottom of all graphs in each figure.}
  \end{figure*}

\begin{figure*}
   \centerline{
 \mbox{{\includegraphics[width=3.50in]{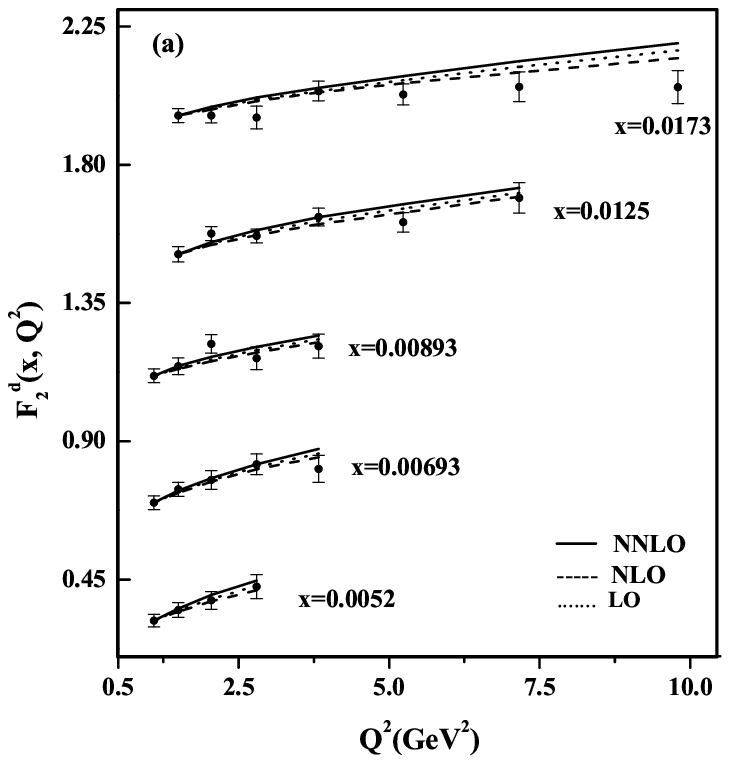}}
{\includegraphics[width=3.50in]{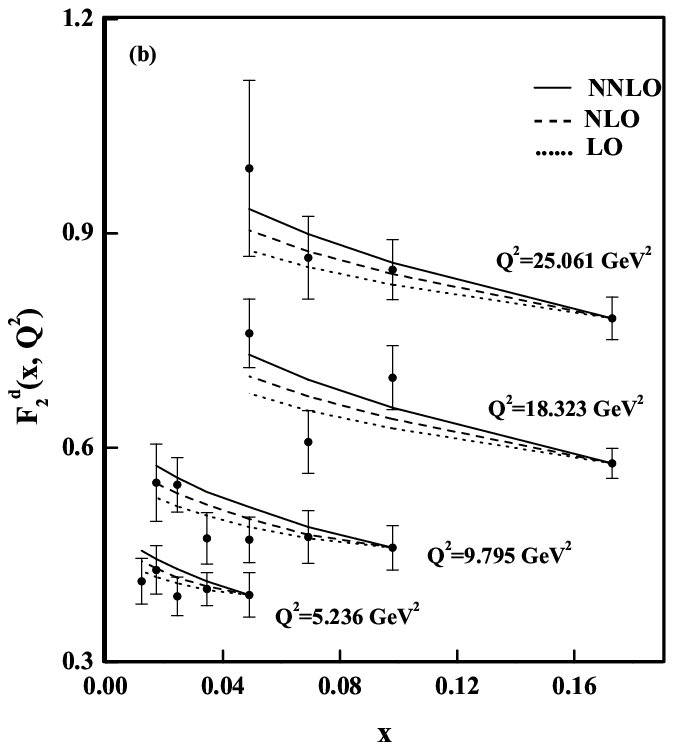}}}
  }
  \caption{E665 data. Dotted lines are LO results, dashed lines are NLO and solid lines are our NNLO results. For clarity, data are scaled up by +0.4i (in Fig.(a)) and +0.1i (in Fig.(b)) (with i = 0,1,2,3) starting from the bottom of all graphs in each figure.}
  \end{figure*}

\begin{figure*}
   \centerline{
 \mbox{{\includegraphics[width=3.50in]{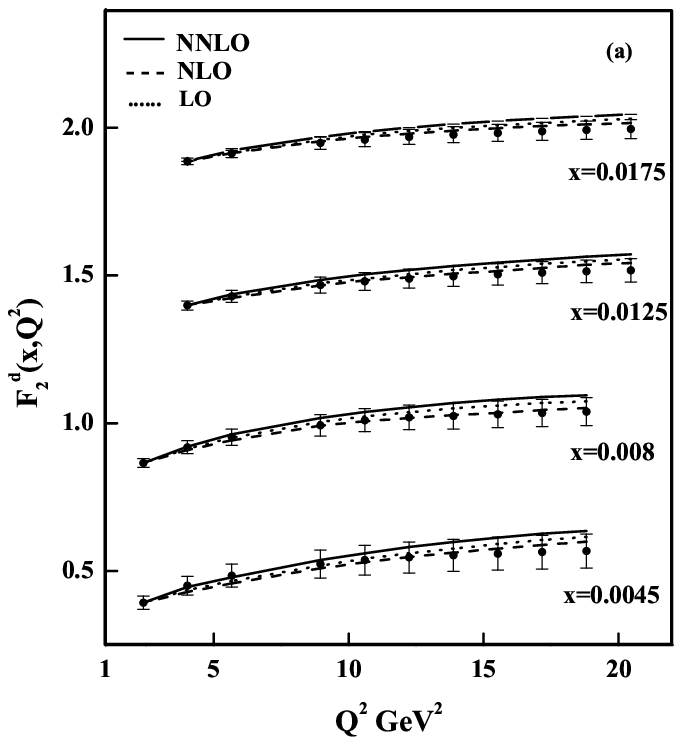}}
{\includegraphics[width=3.50in]{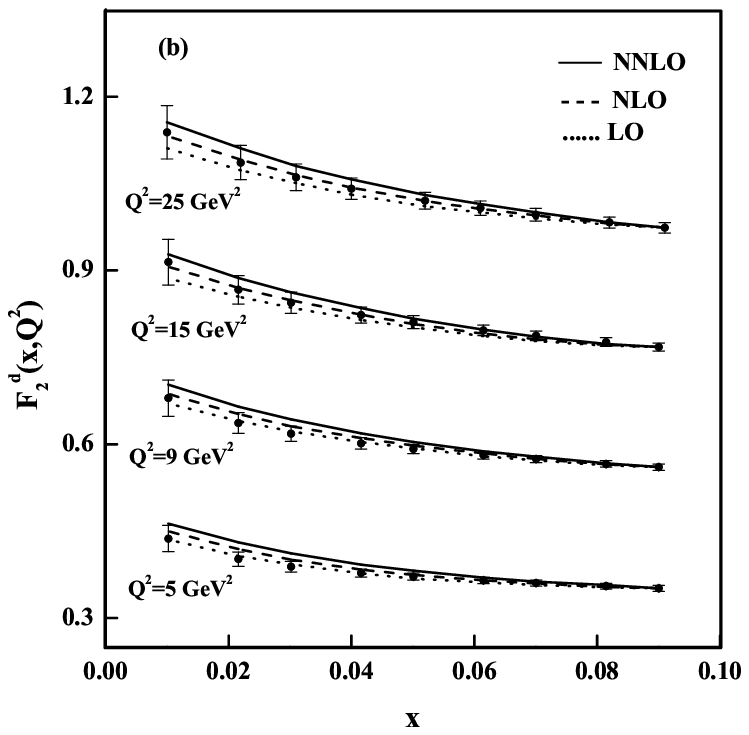}}}
  }
  \caption{NNPDF data. Dotted lines are LO results, dashed lines are NLO and solid lines are our NNLO results. For clarity, data are scaled up by +0.4i (in Fig.(a)) and +0.1i (in Fig.(b)) (with i = 0,1,2,3) starting from the bottom of all graphs in each figure.}
  \end{figure*}

\end{document}